 \documentclass[preprint]{aastex}
 \usepackage{graphicx}
 \usepackage{epsfig}

\shorttitle{Arp~104 Galaxy Pair }
\shortauthors{Gallagher \& Parker}

\begin{document}


\title{Optical Structure and Evolution of the Arp~104 Interacting Galaxy System}


\author{John S. Gallagher III\altaffilmark{1} and Angela Parker\altaffilmark{2}
}

\altaffiltext{1}{Email: jsg@astro.wisc.edu}
\altaffiltext{2}{Current address: Indiana University Astronomy Dept., Swain Hall West 319,  727 East Third Street, Bloomington, IN 47405-7105; parker@astro.indiana.edu}


\begin{abstract}
Arp~104 is a pair of luminous interacting galaxies consisting of NGC~5216,  an elliptical, and NGC~5218, a disturbed disk galaxy and joined by a stellar bridge.  We  obtained optical imaging to support photometric and color studies of the system. NGC~5216 lies on the red sequence, while the unusual distribution of stellar population properties in combination with intense central star formation in a dusty region result in NGC~5218 being a nearby example of an intermediate color (green valley) system.  The stellar bridge has remarkably uniform optical surface brightness, with colors consistent with its stars coming from the outskirts of NGC~5218, but is relatively gas-poor while the northern tidal tail is rich in HI. While both galaxies contain shells, the shell structures  in NGC~5218 are pronounced, and some appear to be associated with counter-rotating gas.  This combination of features suggests that Arp~104 could be the product of distinct multiple interactions in a small galaxy group, possibly resulting from a hierarchical merging process, and likely leading to the birth of a relatively massive and isolated early-type galaxy. 
\end{abstract}

{\em Astrophysical Journal, In Press, August 2010}





\section{Introduction}

Interactions play a major role in the evolution of galaxies, from the hierarchical
growth of small systems at high redshifts to a range of mergers in the current 
universe (e.g., Schweizer 2000).  Even though models have been describing the nature of gravitational 
interactions between galaxies for more than 50 years (e.g., Holmberg 1941, Toomre \& 
Toomre 1972), 
much remains to be learned. For example, debate continues over the nature of 
the interactions even in such well studied systems as M51, where models have been proposed involving either single or multiple close passages of the companion (Toomre 1978, Salo \& Laurakainen 2000, Wahde \& Donner 2001, Dobbs et al. 2009). 


Observational studies of interacting galaxies therefore continue to add value 
in providing empirical descriptions of how perturbed galaxies behave, in the 
process yielding better insights into their structures.  We chose the Arp~104 
system for an optical imaging study based on curiosity concerning its apparently 
extremely smooth optically visible tidal bridge (see Schombert et al. 1990; hereafter SWS90,
Smith et al. 2007),  which led to its inclusion in Arp's 
(1966) {\it Atlas of Peculiar Galaxies}. This primarily stellar bridge connects 
 NGC~5216, the E galaxy southern member of the pair, to the northern disturbed 
 disk galaxy, NGC~5218, over a projected span of 
38~kpc based on an assumed distance of 42~Mpc (0.20~kpc~arcsec$^{-1}$) to the system. 

Beyond its optically visible features, Arp~104 also has several intriguing physical characteristics, many of which are discussed by Roche (2006), Cullen et al. (2007; hereafter C07), and Olsson et al. (2007).  Most of the gas in the system is associated with NGC~5218 and its tidal debris. From their multiwavelength study, C07 infer a molecular gas mass of  $7 \times 10^9$~M$_{\odot}$ in NGC~5218, or more than double the HI mass of $2.8 \times 10^9$~M$_{\odot}$ for the entire Arp~104 system (see also Olsson et al. 2007).  NGC~5218 thus  is well fueled to support its current episode of intense star formation. Smith et al. (2007) further show that the thermal infrared emission is strong and centrally concentrated  in NGC~5218, while thermal infrared is much weaker but also centrally located in NGC~5216.

This study, based on optical photometric imaging from  
the WIYN Observatory 0.9-m and 3.5-m telescopes, focuses on the nature of the interactions 
responsible for Arp~104. We summarize our 
observations in \S2 and present the results in \S3. The astrophysical implications of 
these data are discussed in the context of the interaction and distribution of 
star formation in NGC~5218 in \S4, and 
a summary is provided in \S5.

\section{Observations and Data Analysis} 

\subsection{WIYN 0.9-m}
 Imaging observations were made with the WIYN 0.9m telescope located at the Kitt Peak National Observatory in Arizona on the nights of April 3-9, 2008.  We used the S2KB CCD with an image area of 2048 x 2048 pixels of size 0.60 arcsec creating a field of view of ~20.5 arcmin.
We observed with two standard Harris filters, B and R.  Flat fields and bias exposures were taken at the beginning of each night.   For each of our objects, we took three exposures with each filter, 3 $\times$ 900 seconds in B and 3 $\times$ 600 seconds in R.  Each exposure was dithered from the original position so that CCD flat field deviations could be corrected for in the combining process. The seeing disk full width at half maximum was 1.6~arcsec as estimated from stellar profiles, corresponding to a linear scale of 0.3~kpc. The section of the combined R-band image containing Arp~104 is shown in Figure~1.  

The lower panel of  Figure~1  presents a deeper stretch of the image with a contour located at the 2.5~$\sigma$ detection limit per seeing resolution element of $\mu_r =$25.5~mag~arcsec$^{-2}$.  The signal-to-noise ratio in measurements of features with larger extent than the seeing disk scale with solid angle, $ S/N \propto (\Delta \Omega_{object} / \Delta \Omega_{seeing})^{0.5}$, for the regime where flat field imperfections and scattered light do not introduce additional noise. Measurement precision can improve for objects subtending larger solid angles until flat field errors or other effects dominate, and thus no single detection limit exists for surface brightness studies of resolved objects (e.g., Matthews \& Gallagher 1997). However, the contour in Figure~1 provides a conservative limit to the reliability of our 0.9-m data.

\subsection{WIYN 3.5-m}

\setcounter{footnote}{2}
Arp~104 also was imaged with the WIYN 3.5-m telescope\footnote{The WIYN Observatory is a joint facility of the University of Wisconsin-Madison, Indiana University, Yale University, and the
National Optical Astronomy Observatories.} equipped with the MiniMosaic camera on March 19, 2009 in the R-band, and a narrow band H$\alpha$ filter. The seeing was mediocre, $\approx$1.2~arcsec. Exposure times for these two filters in the resulting combined images were 1800~s, and 5400~s, respectively. These data were not photometrically calibrated, as they were designed to show the inner structures of NGC~5216 and NGC~5218 and detect H{\rm II} region emission.  We also used these images to check for candidate young stellar complexes in the tidal bridge. Due to the gap in the MiniMosaic camera, our coverage of the tidal bridge is incomplete, but no obvious candidate young stellar concentrations with high luminosity were found. 

\subsection{Data Reduction}
Data were reduced using normal tasks in IRAF\footnote{IRAF is distributed by the National Optical Astronomy Observatory, which is operated by the Association of Universities for Research in Astronomy (AURA) under cooperative agreement with the National Science Foundation.} to remove the bias, mask bad pixels, and flat field using dome flats.  Cosmic rays were removed with the lacosmic routine (van Dokkum 2001), and frames were aligned and stacked without additional geometric correction .

The photometric calibration from standard stars observed that night with the 0.9-m telescope proved unreliable.  We therefore transformed the 0.9-m images to the SDSS {\it g} and {\it r} AB magnitude system using stars in the field of view.  The resulting transformations are linear in {\it r} with a small color term. The systematic transformation errors are estimated to be $\leq$0.07 mag in {\it r} and $\leq$0.05 mag in {\it g} $-$ {\it r} from the scatter around the fit.   

The IRAF tasks apphot and ellipse yielded aperture photometry and isophote fitting on the calibrated 0.9-m images and are illustrated in Figures~3 and 5.  Determining colors of the tidal bridge is difficult due to its low surface brightness. However, measurements of the bridge surface brightness show that it is remarkably uniform, varying by $\leq$10\% along the ridgeline between the two galaxies.   Colors of the bridge therefore could be determined by using rectangular apertures along the ridgeline.  A g$-$r color map is shown in Figure~2, and photometric measurements are combined with a summary of other basic information in Table 1.  A check of our g-band photometry transformed to B following Jordi et al. (2006) against the B-magnitude listed in the NASA Extragalactic Database (NED) yields agreement at the 0.1 mag level.  Our results also are generally consistent with those of SWS90 in that we find the bridge to be moderately blue, although we derive somewhat brighter magnitudes for the two main galaxies. We are unable to confirm the reality of the outer blue regions which they found around NGC~5216 and the bridge.

The objectives of the WIYN 3.5-m imaging were to observe HII regions and map smaller angular scale structures.  The central parts of both galaxies are illustrated in Figure~3 along with narrow band images of NGC~5218 made by subtracting an appropriately scaled R-band image as an off-band.  Intense H$\alpha +$[NII] line emission is present in both nuclei (Zhao et al. 2006), and partially accounts for their red g-r colors. Dust lanes also are obvious in the central regions of NGC~5218, consistent with the presence of a larger central concentration of molecular gas and associated strong radio continuum source (C07, Olsson et al. 2007).


\section{Results}

\subsection{NGC~5216}

The structural classification of NGC~5216 as an elliptical galaxy is supported by its mean radial brightness profile which is well fit by an $R^{1/4}$ law from a radii of  3 to $\approx$30~arcsec, i.e. to about 6~kpc (Figure~4 where an R$^{1/4}$ profile will be a line).  Beyond this the presence of shells and the tidal tail produce deviations in the profile, even though the overall trend remains close to an R$^{1/4}$ intensity scaling out to the last reliable data point located slightly beyond r$=$18~kpc. The surface photometry was carried out using the ellipse task in the IRAF package stsdas, with the center position held fixed and the position angle free to be adjusted. This approach also was used to derive the radial brightness distribution of NGC~5218 presented below. Beyond the possibly dusty nucleus with its emission lines, the  {\it g}$-${\it r} color of NGC~5216 also is consistent with a typical elliptical galaxy and it then is not surprising to find that this system lies on the red sequence.  Its LINER nuclear spectrum (Zhao et al. 2006) also is a relatively common feature in an elliptical galaxy.

However, for radii of  R$>$6~kpc, the brightness profile of NGC~5216 flattens, and our results are consistent with those of Roche (2006). In this region shells exist which can be seen in the WIYN 3.5-m images (Figure~5) along with the presence of the southern tidal tail that extends across the galaxy (Figure~3).  The tidal tail is the only region with confirmed recent star formation as indicated by its blue colors.  The outer shells in NGC~5216 unfortunately are too faint, $\mu_r \approx$25~mag~arcsec$^{-2}$, for us to obtain reliable color measurements from our data, although such measurements would be useful in constraining when the shells formed in NGC~5216 (e.g., McGaugh \& Bothun 1990).

\subsection{NGC~5218}

NGC~5218, the northern member of the Arp~104 pair, is well known to have a disturbed structure.  Its fundamental characteristics as a disk galaxy, however, are clear from the optical structure and from results of the radial brightness profile determined from isophotal surface photometry shown in Figure~6 as well as from surface photometry in Roche (2006).  A simple interpretation of Figure~6 is to consider it as indicating the presence of two approximately exponential disks.  The inner disk then has a scale length of $\approx$9~arcsec (2~kpc) while the outer disk scale length increases at radii $>$75 arcsec where the luminosity is increasingly dominated by the tidal debris, which have a shallower mean radial intensity profile than that of the outer disk. 

A  high surface brightness central linear feature with a length of $\sim$15~arcsec (3~kpc), possibly a bar as suggested by C07 and the kinematics found by Olsson et al. (2007), or an edge-on nuclear disk, is seen at PA$\approx$95$^{\circ}$. Beyond the central linear feature is an `inner disk' with major axis at PA$\approx$60$^{\circ}$ and with a radius of 7~kpc.  Optical indications for dust are seen in this region and become prominent in the center of the galaxy, consistent with the intense central far infrared emission observed by Smith et al. (2007) and the presence of substantial molecular gas concentrations (Cullen et al. 2007, Olsson et al. 2007).  Dust therefore can be expected to influence optical colors in the inner parts of NGC~5218, but there are no indications for substantial amounts of dust in the outer disk or tidal material. Olsson et al. further conclude that the nucleus appears to be a starburst-driven LINER based on its radio continuum and molecular line properties.

The inner disk  is encompassed by an `outer disk' with major axis PA$\approx$90$^{\circ}$ that extends to R$\approx$15~kpc. Figure~2 shows that the two main disks also have different colors, with the outer structures being bluest, with colors indicating star formation occurred throughout much of the galaxy at significant levels in the past 10$^9$~yr.  Roche (2006) also found spectroscopic evidence for moderately young stars in NGC~5218 from his integrated spectra which displayed strong H-Balmer absorption lines from an intermediate age stellar component. While the integrated color and absolute magnitude of NGC~5218 place it in the ``green valley", this is not necessarily a proper representation of the global stellar population in this galaxy.  Instead the integrated color reflects a wide range of  local conditions, extending from bluest colors in the outer shell regions where relative star formation rates currently are low to red colors in the dusty central high star formation rate zone.   

The locations of very recent star formation (last $\approx$10~ Myr) traced by the H$\alpha +$[NII] emission support this picture. The majority ($\sim$60\%) of the H$\alpha +$[NII]  flux  occurs in and around the central linear structure, which also is the location of the luminous thermal infrared emission (Smith et al. 2007). The remainder is distributed primarily along the rim of the inner disk, with a few HII regions detected in the outer disk.  The presence of HII regions indicates that the reddening of g$-$r color toward the center of NGC~5218  is  associated with the distribution of dust (see Figure~3 and the far infrared maps of Smith et al. (2007)) rather than an absence of young stellar populations. In particular the very red  nuclear color that we observe apparently is fostered by a combination of dust and intense H$\alpha +$[NII] emission, and follows the tendency for dusty central star formation seen in samples of green valley galaxies (Cowie \& Barger 2008, Brammer et al. 2009).

The impression that the two main disks are tilted with respect to one another is supported by the kinematic data presented in C07. They note that the inner velocity field as measured from the CO 1-0 transition appears to rotate counter to that defined by the more radially extended HI (Figure~3).  This can be understood as result of a strongly warped disk component that changes in angle with radius (e.g. Sparke et al. 2009).  We therefore conclude that the disk system in NGC~5218 is unlikely to be coplanar, but instead shows the large range in angular momentum vector direction that is typical of galaxies which have experienced recent mergers (Barnes 2002).   The combination of structures and kinematics in NGC~5218 is reminiscent of the out-of-plane features seen in polar ring galaxies, such as NGC~3718 (Sparke et al. 2009), and suggests that NGC~5218 is the product of a relatively recent merger. 

As a result of this event, the optical structure of NGC~5218  is still evolving. In particular the central concentration of star formation along with its high dust content compromise our ability to determine the ultimate bulge-to-disk ratio of the merger remnant from its current optical structure (see Lotz et al. 2008). Thus we cannot tell from optical surface photometry alone if NGC~5218 is a product of a major merger of near equal mass galaxies, or due to some other class of minor merger event.  Measurement of the stellar velocity field will be helpful in revealing the inner structure of NGC5218 and thus clarifying this issue. For example, kinematic data, especially if obtained in the near-infrared, can reveal whether the central linear feature is likely to be a bar observed at moderate inclination or a possibly a combination of bar and highly inclined central disk (c.f. Olsson et al. 2007), and whether a significant spheroidal bulge is present.

Following the discussion of C07 and using the results from Surace et al. (2004), NGC~5218 has SFR $\approx$5-7~M$_{\odot}$~yr$^{-1}$. The gas exhaustion time scale with the current SFR is $\sim$1.3-1.9~Gyr, or longer than the likely time for the two systems to merge (see also C07). 

\subsection{Stellar Tidal Bridge and Tails}

The stellar bridge between galaxies is well resolved with a the width of $\sim$10~arcsec (2.2~kpc), and {\it r}-band luminosity of only  $\sim$2\% of that of NGC~5218. It also  
is remarkably constant in surface brightness. Combining the 0.9-m telescope B- and R-band images in an unweighted average to gain signal-to-noise, the bridge surface brightness determined in 5.4$\times$5.4~arcsec$^2$ regions remains constant to $< \pm$10\%. The only indication of small scale structure is a  marginally detected decrease in surface brightness by 
$\sim$20\% in the vicinity of the  HI emission peak mapped by C07.   The structure perpendicular to the long axis of the bridge is similarly regular, with a slow rise in intensity from the east and a sharp drop off to the west after the intensity peak.  The form of the spatial gradient in surface brightness perpendicular to the long axis of the bridge is consistent with its origin in a prograde interaction between NGC~5216 and NGC~5218 (Oh et al. 2008), in agreement with the basic features of the C07 NGC~5216-NGC~5218 interaction model.

We estimated the minimum stellar mass density along the ridge of the tidal bridge by adopting a single age GALEV stellar population model (Kotulla et al. 2009) with a Kroupa-IMF and slightly sub-solar metallicity appropriate to an outer galactic disk. The {\it minimum} stellar mass was found by assuming the bridge consists of a 1.1~Gyr age simple stellar population which would have {\it g}$-${\it r} $\approx$ 0.5. Note that a more realistic model is likely to contain older stars and thus have a higher mass-to-light ratio, but even with only a {\it g}$-${\it r} color to constrain the stellar population parameters, this simple model provides an approximate lower limit to the stellar mass density.  This match gives $\Sigma_* \approx $20~M$_{\odot}$~pc$^{-2}$, and a mean stellar density of $\rho_* \sim$0.01~($\Delta /2$~kpc)~M$_{\odot}$~pc$^{-3}$ where $\Delta$ is the depth of the bridge along the line of sight.   These results indicate that the bridge is not rich in HI.   The mean value of N(HI)$>$10$^{21}$~cm$^{-2}$ required to equal the stellar density readily exceeds the C07 measured peak value of N(HI)$\approx$10$^{20}$~cm$^{-2}$.  We therefore conclude that $\Sigma_{HI} / \Sigma_* \leq$ 0.1 along the bridge. 

Optical bridges between galaxies originate from stars which tides strip from outer galactic disks of interacting galaxies (e.g., Toomre \& Toomre 1972, Oh et al. 2008). The optical color of the Arp~104 bridge is constant at g$-$r$=$0.55$\pm$0.1, and, in agreement with SWS90, the bridge color is  consistent with its stellar populations originating in the outer disk of NGC~5218 (see Table~1 and Roche (2006)). We find no evidence for substantial levels of ongoing star formation along the bridge between galaxies, but cannot exclude the presence of small star forming regions, such as those seen in the M81 tidal bridge (de Mello et al. 2008). 

The northern and southern optical tidal tails, are much shorter than the bridge, with less regular structures. Based on the {\it g}$-${\it r} color, significant young stellar populations with ages of $<$10$^9$~yr are found in the northern tidal tail, and, perhaps surprisingly, also in its southern counterpart that extends from NGC~5216 .  The northern optical tidal feature is located at the base of the long HI tidal tail that extends to the northeast of NGC~5218, most of which is not detected in the optical. 

Arp~104 therefore shows a highly asymmetric set of tidal features. The majority of the HI gas lies in NGC~5218 and the northern tidal tail, while most of the tidally stripped stars are in the inter-galaxy bridge. 

\subsection{The Environment}

Arp~104 is the dominant member of a small group of galaxies, LGG~854 (Garcia 1993).  The other   cataloged member of LGG~854 is NGC~5216A, an edge-on late-type disk galaxy at a projected distance of $R_P =$580~kpc. A search of NED also yields UGC~8436 at $R_P =$ 680~kpc and UGC~8491 at $R_P =$230~kpc, both moderate luminosity late-type galaxies, and possible group members.  A search of the 0.9-m images yielded only two LSB objects, but given the distance to Arp~104 and image quality of the 0.9-m data, these cannot be reliably classified. However, we find no luminous dwarfs with M$_r <-$15 within the 240~kpc field covered by the 0.9-m telescope images. 

\section{Discussion}

The combination of our optical measurements of the stellar structure and colors of Arp~104 and multi-wavelength studies of Roche (2006) and C07 indicate that Arp~104 is the product of an interesting sequence of astrophysical processes:
\begin{itemize}

\item As described by C07, a prograde close passage between the disky NGC~5218 galaxy and the elliptical NGC~5216 occurred about 300~Myr in the past.  The smooth stellar tidal bridge is a product of that interaction and comparisons with models show the two galaxies are likely to be separating. The shearing along the tidal bridge as the two galaxies separate, along with its low stellar density, may explain the near constant surface brightness, since gravitational instabilities will have difficulty in growing under these conditions.  As discussed by Oh et al. (2008), the surface brightness of tidal bridges increase with interaction strength, so the low brightness of the Arp~104 bridge then suggests that NGC~5218 did not make an extremely close passage by NGC~5216. 

\item The tidal tails in the system include features with HI gas and  younger stellar populations extending to the southwest from NGC~5216 with similar conditions present at the base of primary HI tail running to the north and east of NGC~5218.  The northern tidal tail, however, is optically faint and therefore stands in contrast to the bridge joining the two galaxies, where stars dominate. The origin of this asymmetry is not immediately obvious in the context of a purely binary galaxy interaction model. Our color estimates and those of SWS90 for the tidal bridge are consistent with its stars coming from NGC~5218, and are insufficiently blue for the bridge to consist primarily of stars formed during the interaction. 

\item NGC~5218 contains stellar shells with the bluest optical colors in the galaxy, showing that they hosted significant star formation within the past $\sim$1~Gyr.  These features are not reproduced by the interacting galaxy model of C07.  Stellar shells usually are associated with mergers involving disk galaxies (e.g., Schweizer \& Seitzer 1988), which raises the question as to whether NGC~5218 experienced a merger before interacting with NGC~5216.  Two possibilities are that this material came from an interaction with NGC~5216, or involved a minor merger of NGC~5218 with a third galaxy. Given the red sequence colors, elliptical galaxy structure without an obvious disk, and apparent lack of a cool interstellar medium in NGC~5216, we favor the recent minor merger option for the origin of shells in NGC~5218.  A previous merger could have broken the axial symmetry of NGC~5218, which could lead to the currently observed structural differences in the tidal tails.  Modeling is needed to explore this possibility. 

\item  The distribution of `disk' HII regions in two ellipsoidal rings inclined to each other also suggests that NGC~5218 experienced a significant external perturbation.  While a previous encounter with NGC~5216 might explain the situation, the counter rotation observed in the $^{12}$CO line by C07 is more readily achieved by material associated with a previous merger (Bureau \& Chung 2006, Vergani et al. 2007, Crocker et al. 2009 and references therein). 

\item The integrated colors of NGC~5218 place it in the green valley, between the blue plume and red sequence galaxies. In agreement with other recent studies (e.g., Cowie \& Barger 2008, Brammer et al. 2009), this color reflects a combination of blue colors from young but aging stellar populations and a dusty central region.  If a future merger were not a factor, then NGC~5218 contains sufficient gas to fuel star formation at a rate that would allow it to remain in something close to its current state for up to $\sim$1~Gyr.  

\item NGC~5216 also displays faint outer shell structures.  Their presence complicates the interpretation of this system.  Are these remnants of yet another previous interaction? Or could these features result from material captured from NGC5218? If so we would expect them to have blue optical colors, typical of the outer regions of NGC5218. Unfortunately we were not able to make the necessary measurements from our data. 

\item Both galaxies host energetic processes in their nuclei.  NGC~5216 has a LINER optical spectrum (Zhao et al. 2006) and the H$\alpha +$[NII] emission probably is a major source of the very red nuclear colors that we observe. Optically visible star formation, the radio continuum (C07, Olsson et al. 2007) and thermal infrared emission (Smith et al. 2007) is strongly concentrated within the central linear feature of NGC~5218, which produces most of its H$\alpha +$[NII] emission line luminosity. Since this region also contains obvious dust lanes and and a dense concentration of molecular gas (Olsson et al. 2007), this is a lower bound on the H$\alpha$ luminosity fraction, which will increase when corrected for internal extinction. The multi-wavelength data are consistent with the presence of a nuclear starburst and possible LINER in NGC~5218. 

\item Multiple mergers have been suggested to be factors in early-type galaxies involved in interactions in galaxy groups (e.g., P\'{e}rez Grana et al. 2009). Thus Arp~104 potentially illustrates a significant evolutionary pathway for the production of early-type galaxies in small groups.  In this case we find products of possible merging sequences where NGC~5218 combined with another system a few internal dynamical times ago and probably will merge with NGC~5216 a few internal dynamical times in the future. This process contrasts with the possibility of rapid multiple mergers, e.g., in which more than two members of a compact galaxy group might combine to form one galaxy on approximately a compact region crossing time or few internal galactic dynamical times (e.g., Rubin et al. 1990, Amram et al. 2007).

\end{itemize}

Based on standard galaxy interaction models, such as that in C07, the eventual fate of Arp~104 will be to merge into a single galaxy, as also suggested by Roche (2006).  Whether a sufficient reservoir of gas with high specific angular momentum will remain to allow a star-forming disk to exist in the resulting merged system is unclear.  While the current supply of gas ($\approx$10$^{10}$~M$_{\odot}$) is adequate to make a moderately luminous stellar disk, its  eventual state depends on the properties of gas in the tidal tails in combination  with the effectiveness of processes in  funneling gas into the central region where high densities promote dissipation and star formation and can lead to building a bulge (cf. Hopkins et al. 2008).  

Although standard spiral-spiral merger models predict that only a small disk is likely to remain once a merged system settles into its final form over $\sim$0.3-1~Gyr (e.g., Lotz et al. 2008, Hopkins et al. 2009b), this conclusion may not apply to major spiral-elliptical mergers, such as that in the future for Arp~104 (Di Matteo et al. 2008). Furthermore, a sequence of  mergers closely spaced in time, as we are suggesting  for Arp~104, could affect the angular momentum of gas in the tidal tails, and therefore the structure of the final merged system.  Thus the galaxy resulting from the NGC~5216-NGC~5218 major merger might have either an approximately Sa-type morphology if some of the NGC~5218 disk is retained and renewed through post-merger star formation, or an S0/E structure if dissipation dominates and gas primarily collects in the central region of the merged galaxy (Bekki 2001, Hopkins et al. 2009a). 

As the group containing Arp~104 at most contains a few widely scattered and much less luminous  members, this system could be headed towards becoming what appears to be an isolated elliptical galaxy (Aars et al. 2001).  However, it will have achieved this state through successive mergers, and thus the isolation would be a product of environmental "nurture" in a small galaxy group, rather than "nature" in the form of initial conditions leading to the formation of a relatively massive early-type galaxy at high redshift (e.g. Bournaud et al. 2007).

\section{Summary}

Our study of the optical structure of Arp~104 reveals features  which illuminate the evolutionary state of this system. The presence of shells in both members of the pair and apparent counter-rotation in NGC~5218 suggest that they both have been influenced by interactions. These features possibly could result from earlier encounters between the two galaxies, or, given the angular momentum misalignment in NGC~5218, in our view more likely from its involvement in an earlier merger. This also would have distorted the system and therefore might be the origin of the asymmetry in the tidal bridges, with HI dominating to the north and stars between NGC~5216 and NGC~5218. If this idea is correct, then the structure of the ultimate merger  of the Arp~104 system also could be affected, e.g., if infalling tidal material were to have gained sufficient angular momentum to avoid being incorporated into a bulge and instead aiding the survival of a disk component.

The disturbed structure of NGC~5218 suggests interactions moved gas inwards, leading to star formation becoming centrally concentrated in a dusty, molecular gas-rich nuclear starburst zone. The integrated optical colors place this galaxy in the green valley, while NGC~5216 is on the red sequence. However, consistent with other recent statistical studies, the green valley color of NGC~5218 stems from a wide range of recent star formation histories and does not properly represent the current state of the galaxy, which supports intense star formation in its dusty central regions. 

While deeper imaging would be useful, the lack of bright dwarfs or tidal debris differing from those readily associated with the binary pair Arp~104 appears inconsistent with expectations if this system were the late stage product of dynamical timescale merging in compact galaxy group. We therefore favor a hierarchical merger model where interactions first led to NGC~5216 and NGC~5218 dominating their galaxy group, with the future product being a merger into a single luminous Sa or earlier type galaxy that we now catch in the process of formation.

\acknowledgments
This first phase of this project comprised A. Parker's senior thesis in Astronomy at the University of Wisconsin-Madison.  We thank the WIYN Observatory teams, especially Hillary Mathis of the WIYN 0.9-m Obervatory, for making it possible to obtain the observations reported here, Kate Dellenbusch for her help with the WIYN 0.9-m observations, and an anonymous referee for useful comments and for pointing out the unpublished very interesting study of Arp~104 by N. Roche.  JSG benefited from discussions with Susanne Aalto and Evert Olsson during a visit to the Onsala Space Observatory. He also thanks the University of Canterbury, Christchurch, New Zealand for the opportunity to hold an Erskine Visiting Fellowship during the revisions of this manuscript. This research was supported in part by the National Science Foundation through grant AST-0708967 to the University of Wisconsin and has made use of the NASA/IPAC Extragalactic Database (NED) which is operated by the Jet Propulsion Laboratory, California Institute of Technology, under contract with the National Aeronautics and Space Administration.

\begin{center}
\begin{table}
\caption{Photometric Measurements of Arp104}
\begin{tabular}{ccllc}
\tableline \tableline
Object & Region & {\it r} or $\mu_r$ & {\it g} $-$ {\it r} & Comment \\
\tableline
NGC~5216  & integrated &  12.44$\pm$0.05 \tablenotemark{a} & 0.80$\pm$0.05 &  M$_r = -$20.7 \tablenotemark{b} \\
                               & envelope  &   & 0.83$\pm$0.03 \\
NGC~5218 & integrated  & 11.96$\pm$0.1 & 0.60$\pm$0.05 &  M$_r = -$21.2 \\
                      & nuclear zone    &                        &   0.78$\pm$0.03    & Average of 2 sides \\
                      & outer arms &                        &   0.45$\pm$0.05     &     \\
North tidal tail  & Outer, R$\geq$60~arcsec  & $\mu_r \geq $23.8\tablenotemark{c} &    0.54$\pm$0.05       &   \\
Bridge             &			&$\mu_r =$24.2 &   0.55$\pm$0.1        &  Constant $\mu_r$ to $\approx$10\%   \\
South tidal tail &  R=55~arcsec   & $\mu_r \geq $23.6   &   0.50$\pm$0.1        &      \\
\tableline                             
\end{tabular}
\tablenotetext{a}{Errors refer to uncertainties in measurements and do not include 
estimated 0.07 and 0.05 mag \\ systematic errors in transformations to {\it r} and {\it g} $-$ {\it r}, respectively.}
\tablenotetext{b}{Absolute magnitudes in SDSS system based on D$=$42~Mpc and A$_r =$0.05.}
\tablenotetext{c}{Surface brightnesses are in units of r magnitudes~arcsec$^{-2}$; estimated measurement uncertainty is $\pm$0.1~magnitude/arcsec$^{2}$. }
\end{table}
\end{center}

\pagebreak

\begin{figure}
\begin{center}
\leavevmode
\includegraphics[angle=0,scale=0.850]{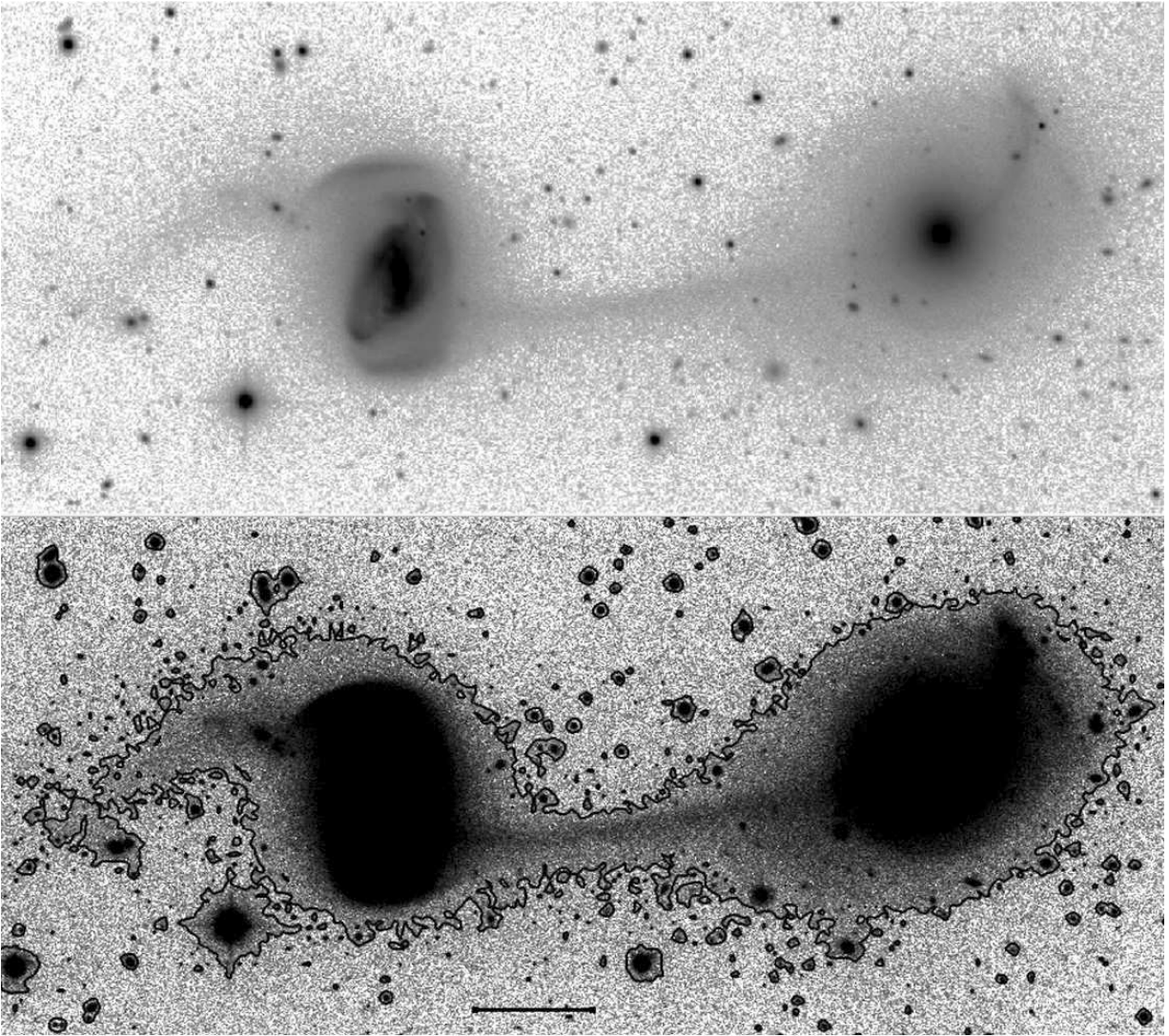} 
\caption{Arp 104 system observed in the R-band with the WIYN 0.9-m telescope; N to left and E is down, so NGC~5218 is to the left and NGC~5216 to the right and the width of each panel in the image is 8.8~arcmin corresponding to $\sim$110~kpc. The lower image shows a deeper stretch with the contour located at the 2.5~$\sigma$ detection level for an object with the 1.6~arcsec seeing disk full width at half maximum, corresponding to 0.3~kpc. The bar in the lower panel is 10~kpc in length. \label{arp104_r1}}
\end{center}
\end{figure}

\begin{figure}
\begin{center}
\includegraphics[angle=0,scale=0.850]{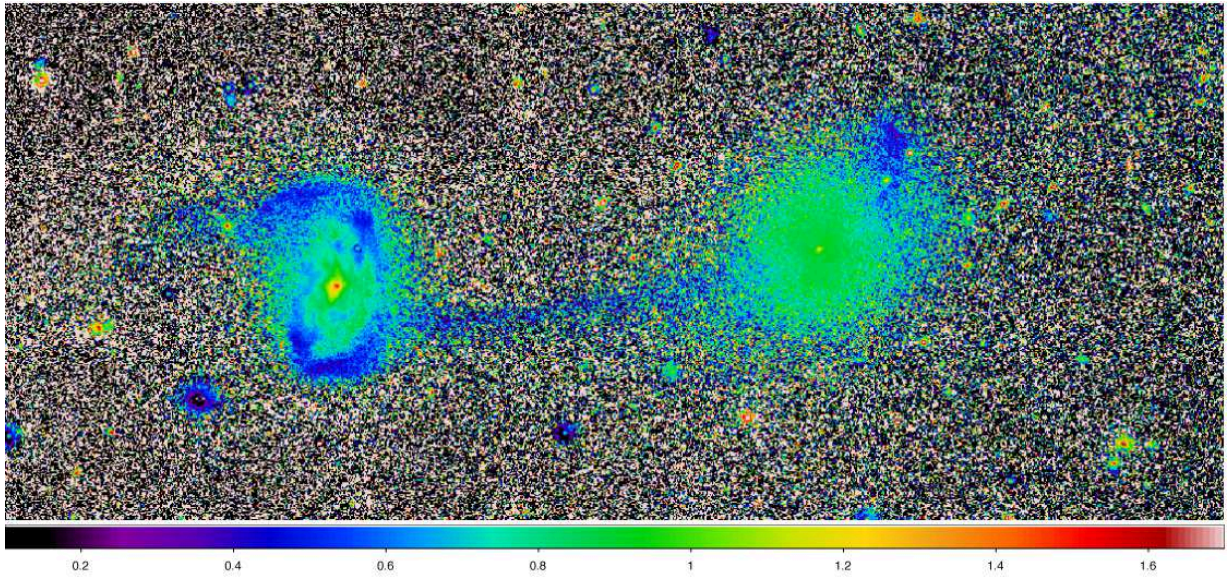}
\caption{Arp 104 system {\it g}$-$ {\it r} color map from observations with the WIYN 0.9-m telescope; N to left and E is down. The field of view is the same as in Figure~1. The central part of the tidal bridge is at the low surface brightness limit for reliable {\it g}$-${\it r} color measurements with these data. \label{arp104cols}}
\end{center}
\end{figure}

\begin{figure}
\begin{center}
\includegraphics[angle=0,scale=0.90]{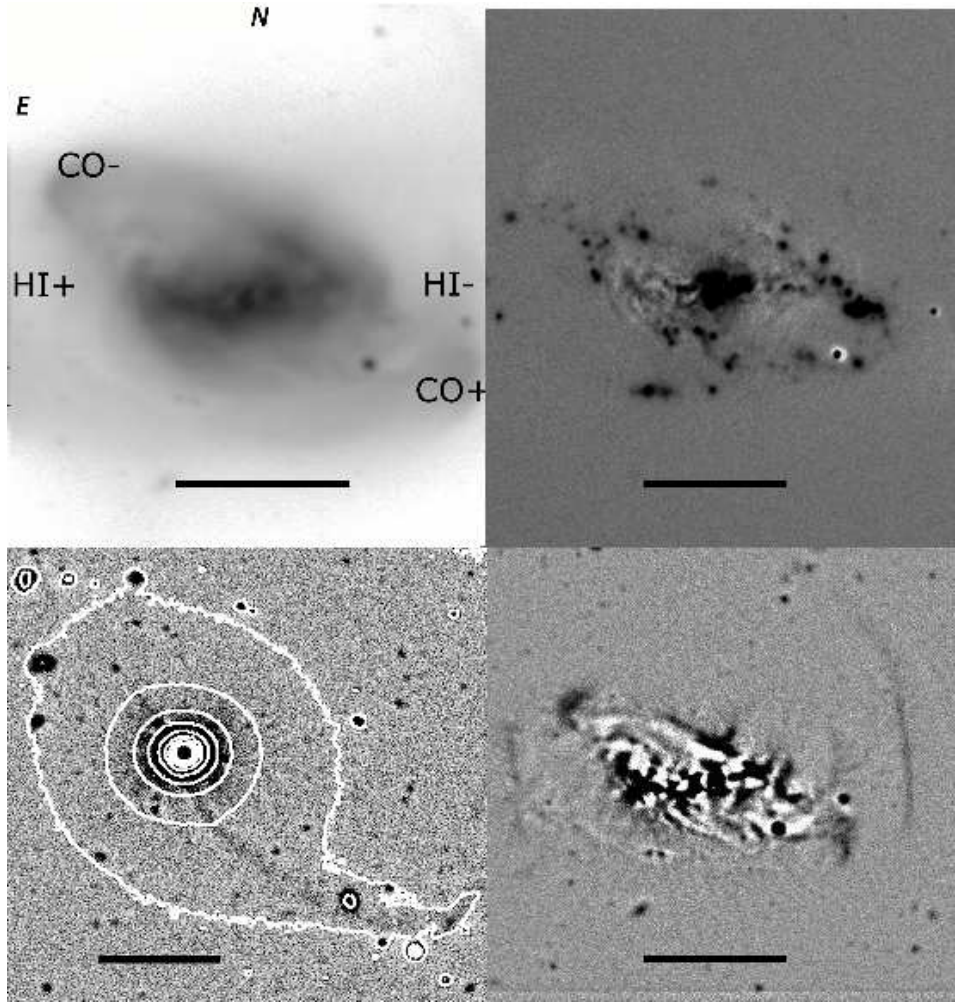}
\caption{Clockwise: NGC~5218 from WIYN 3.5-m images in R-band with velocity directions ($+$ redshifted with respect to central velocity, $-$ blueshifted) from C07 (upper left); H$\alpha +$ [NII] narrow band (upper right); NGC~5218 high pass image showing the dust lanes and outer shells (lower right). NGC~5216 high pass spatially filtered WIYN 3.5-m R-band image with stellar intensity contours showing the southwestern tidal tail extending from the center of NGC~5216 (lower left). \label{arp104_35}}
\end{center}
\end{figure}

\begin{figure}
\begin{center}
\includegraphics[angle=0,scale=0.50]{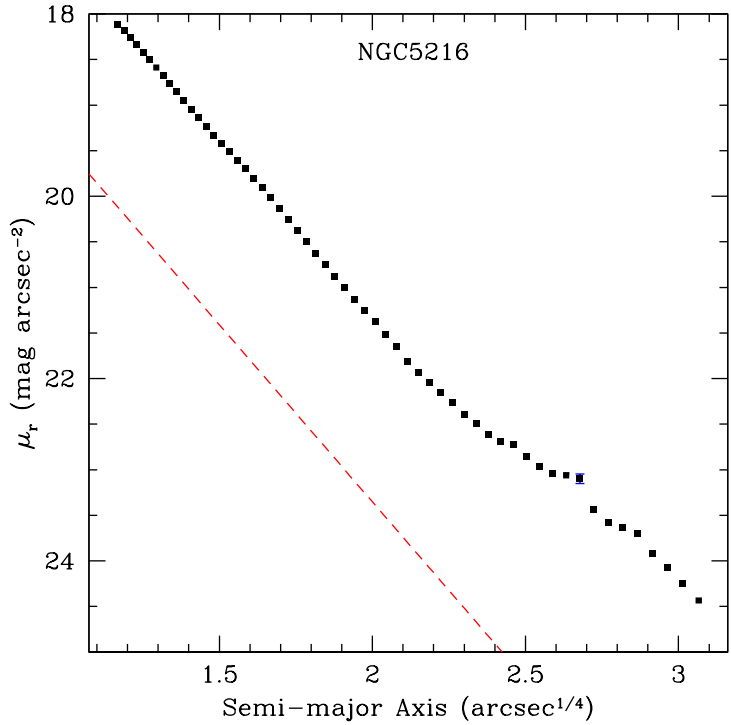}
\caption{Radial surface brightness plot of NGC~5216 from the WIYN 0.9-m image and a line showing the $R^{1/4}$-law profile typical of an elliptical galaxy. The actual data are very close to the R$^{1/4}$ profile in the inner regions, but  display significant deviations at larger radii associated with the outer shells. Error bars from the ellipse fitting task are plotted (in color online for increased visibility) when they exceed 0.05 mag.  R$^{1/4} =$3 arcsec$^{1/4}$ corresponds to r$=$16.5~kpc. \label{n5216_isophot}}
\end{center}
\end{figure}

\begin{figure}
\begin{center}
\includegraphics[angle=0,scale=0.750]{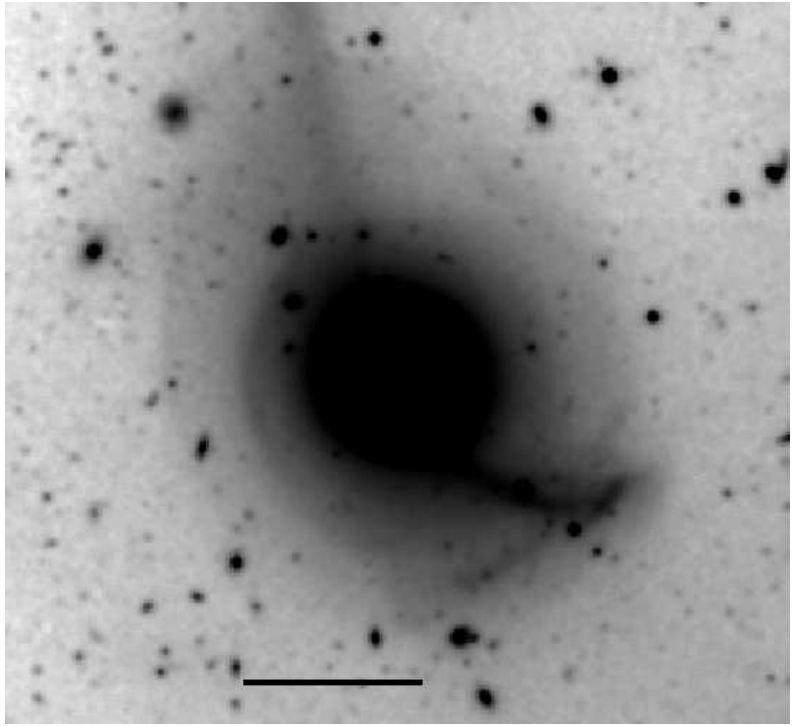}
\caption{Image of NGC~5216 from WIYN 3.5-m R-band imaging showing the southwestern tidal tail and faint outer shells. North is up, east to the left, and the width of the image is 3.7~arcmin corresponding to 45~kpc and the scale bar shows a 10~kpc length for comparison with the lower left panel of Figure~3 . \label{n5216_deepr}}
\end{center}
\end{figure}

\begin{figure}
\begin{center}
\includegraphics[angle=0,scale=0.50]{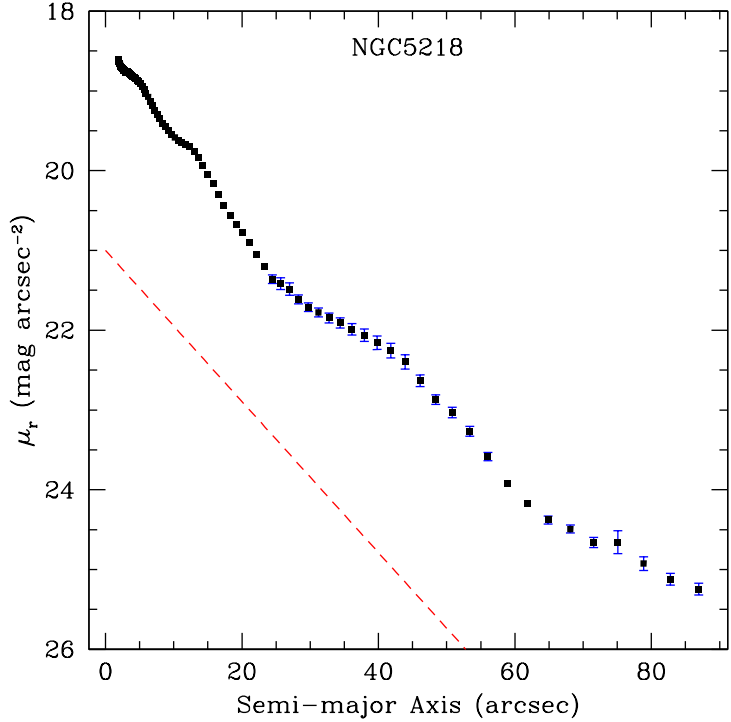}
\caption{Radial surface brightness plot of NGC~5218 from the WIYN 0.9-m image showing the overall exponential brightness profile typical of a galactic disk.  Error bars from the ellipse fitting task are plotted when they exceed 0.05 mag (in color online for increased visibility) and the scale is 200~pc arcsec$^{-1}$. At R$>$75~arcsec the presence of tidal debris make surface photometry of the disk increasingly unreliable. The line shows an exponential disk model with a scale length of 2~kpc and central surface brightness offset by $+$2.4~mag~arcsec$^{-2}$ for visibility from the the mid-disk model fit of $\mu_r(0) =$18.6 mag~arcsec$^{-2}$.  \label{n5216_deepr}}
\end{center}
\end{figure}

\end{document}